# Dynamics of Mixed Dark Energy Domination in Teleparallel Gravity and Phase Space Analysis

## Emre DİL[1], Erdinç KOLAY[2]

[1]Department of Physics, Sinop University, 57000, Korucuk, Sinop-TURKEY
[2]Department of Statistics, Sinop University, 57000, Korucuk, Sinop-TURKEY

**Abstract:** We consider a novel dark energy model to investigate whether it will provide an expanding universe phase. Here we propose a mixed dark energy domination which is constituted by a tachyon, quintessence and phantom scalar fields non-minimally coupled to gravity, in the absence of background dark matter and baryonic matter, in the framework of teleparallel gravity. We perform the phase-space analysis of the model by numerical methods and find the late-time accelerated attractor solutions implying the acceleration phase of universe.

## 1. Introduction

The universe is known to be experiencing an accelerating expansion by astrophysical observations such that, Supernova Ia [1,2], large scale structure [3,4], the baryon acoustic oscillations [5] and cosmic microwave background radiation [6-9].

In order to explain the late-time accelerated expansion of universe, an unknown form of energy, called as dark energy is proposed. This unknown component of energy possesses some interesting properties, for instance, it is not clustered but spread all over the universe and its pressure is negative for driving the current acceleration of the universe. Observations show that the dark energy occupies 70% of our universe.

What is the constitution of the dark energy? One candidate for the answer of this question is the cosmological constant $\Lambda$ having a constant energy density filling the space homogeneously [10-13]. But cosmological constant is not well accepted since the cosmological problem [14] and the age problem [15]. For this reason, many other dark energy models have been proposed instead of the cosmological constant. Other candidates for dark energy constitution are quintessence, phantom and tachyon fields. We can briefly classify the dark energy models in terms of the most powerful quantity of dark energy; its equation of state parameter $\omega_{DE} = p_{DE}/\rho_{DE}$, where $p_{DE}$ and $\rho_{DE}$ are the pressure and energy density of the dark energy, respectively. For cosmological constant boundary $\omega_{DE} = -1$, but for quintessence the parameter $\omega_{DE} \geq -1$, for phantom $\omega_{DE} \leq -1$ and for non-minimally coupled tachyon with gravity both $\omega_{DE} \leq -1$ and $\omega_{DE} \geq -1$ [16-18].

The scenario of $\omega_{DE}$ crosses the cosmological constant boundary is referred as a "Quintom" scenario. The explicit construction of Quintom scenario has a difficulty, due to a no-go theorem. The equation of state parameter $\omega_{DE}$ of a scalar field cannot cross the cosmological constant boundary according to no-go theorem, if the dark energy described by the scalar field is minimally coupled to gravity in Friedmann-Robertson-Walker (FRW) geometry. The requirement for crossing the cosmological constant boundary is that the dark energy should non-minimally coupled to gravity, namely it should interact with the gravity [19-24]. There are also models in which possible coupling between dark energy and dark matter can occur [25,26]. In this paper, we consider a mixed dark energy model constituted by a tachyon, quintessence and phantom scalar fields non-minimally coupled to gravity.

The mixed dark energy model in this study is considered in the framework of teleparallel gravity instead of classical gravity. The teleparallel gravity is the equivalent form of the classical gravity, but in place of torsion-less Levi-Civita connection, curvature-less Weitzenbock one is used. The Lagrangian of teleparallel gravity contains torsion scalar $T$ constructed by the contraction of torsion tensor, in contrast to the Einstein-Hilbert action of classical gravity in which contraction of the curvature tensor $R$ is used. In teleparallel gravity, the dynamical variable is a set of four tetrad fields constructing the bases for the tangent space at each point of space-time [27-29]. The teleparallel gravity Lagrangian with only torsion scalar $T$ corresponds to the matter-dominated universe, namely it does not accelerate. Therefore, to obtain a universe with an accelerating expansion, we can either replace $T$ with a function $f(T)$, the so-called $f(T)$ gravity (teleparallel analogue of $f(R)$ gravity) [30-32], or add an unknown form of energy, so-called dark energy, into the teleparallel gravity Lagrangian that allows also non-minimal coupling between dark energy and gravity to overcome the no-go theorem. The interesting feature of $f(T)$ theories is the existence of second or higher order derivatives in equations. Therefore, we prefer the second choice; adding extra scalar fields of the unknown energy forms as dark energy.

As different dark energy models, interacting teleparallel dark energy studies have been introduced in the literature, for instance Geng *et. al.* [33,34] consider a quintessence scalar field with a non-minimal coupling between quintessence and gravity in the context of teleparallel gravity. The dynamics of this model has been studied in [35-37]. Tachyonic teleparallel dark energy is a generalization of the teleparallel quintessence dark energy by introducing a non-canonical tachyon scalar field in place of the canonical quintessence field [18,38-40].

In this study, the main motivation is that we consider a more general dark energy model including three kinds of dark energy models, instead of taking one dark energy model as in [16,18,38-40]. In order to explain the expansion of universe by adding scalar fields as the dark energy constituents, there has never been assumed a cosmological model including three kinds of dark energy models. We assume a tachyon, quintessence and phantom fields as a mixed dark energy model which is non-minimally coupled to gravity in the framework of teleparallel gravity. We make the dynamical analysis of the model in FRW space-time. Later on, we translate the evolution equations into an autonomous dynamical system. After that the phase-space analysis of the model and the cosmological implications of the critical (or fixed) points of the model will be studied from the stability behavior of the critical points. Finally, we will make a brief summary of the results.

## 2. Dynamics of the model

Our model consists of three scalar fields as the three-component dark energy domination without background dark matter and baryonic matter. These are the canonical scalar field, quintessence $\phi$ and two non-canonical scalar fields, tachyon $\psi$ and phantom $\sigma$ which three of the scalar fields are non-minimally coupled to gravity. Since we consider only the dark energy dominated sector without the matter content of the universe, the action of the mixed teleparallel dark energy with a non-minimal coupling to the gravity can be written as [16,38,39]

$$S = \int d^4 x \, e \left[ \frac{T}{2\kappa^2} + \xi_1 \, f(\psi) T - V_\psi(\psi) \sqrt{1 + g^{\mu\nu} \frac{\partial_\mu \psi \, \partial_\nu \psi}{V_\psi(\psi)}} \right.$$
$$\left. + \xi_2 \, g(\phi) T - \frac{1}{2} g^{\mu\nu} \partial_\mu \phi \, \partial_\nu \phi - V_\phi(\phi) \qquad , \right. \tag{1}$$
$$\left. + \xi_3 \, h(\sigma) T + \frac{1}{2} g^{\mu\nu} \partial_\mu \sigma \, \partial_\nu \sigma - V_\sigma(\sigma) \right]$$

where $e = \det(e^i_\mu) = \sqrt{-g}$ and $e^i_\mu$ are the orthonormal tetrad components, such that

$$g_{\mu\nu} = \eta_{ij} e^i_\mu e^j_\nu, \tag{2}$$

where $i, j$ run over 0, 1, 2, 3 for the tangent space at each point $x^\mu$ of the manifold and $\mu$, $\nu$ take the values 0, 1, 2, 3 and are the coordinate indices of the manifold. While $T$ is the torsion scalar, it is defined as

$$T = \frac{1}{4} T^{\rho\mu\nu} T_{\rho\mu\nu} + \frac{1}{2} T^{\rho\mu\nu} T_{\nu\mu\rho} - T^\rho_{\mu\nu} T_\rho^{\nu\mu}. \tag{3}$$

Here $T^\rho_{\mu\nu}$ is the torsion tensor constructed by the Weitzenbock connection $\Gamma^\rho_{\mu\nu}$, such that [41]

$$T^\rho_{\mu\nu} = \Gamma^\rho_{\nu\mu} - \Gamma^\rho_{\mu\nu} = e^\rho_i (\partial_\mu e^i_\nu - \partial_\nu e^i_\mu). \tag{4}$$

All the information about the gravitational field is contained in the torsion tensor $T^\rho_{\mu\nu}$ in teleparallel gravity. The Lagrangian of the theory is set up according to the conditions of invariance under general coordinate transformations, global Lorentz transformations and the parity operations [42].

Furthermore, $\kappa^2 = 8\pi G$ in (1) and $f(\psi)$, $g(\phi)$ and $h(\sigma)$ are the functions responsible from non-minimal coupling between gravity $T$ and tachyon, quintessence and phantom fields, respectively. $\xi_1$, $\xi_2$ and $\xi_3$ are the dimensionless coupling constants and $V_\psi(\psi)$, $V_\phi(\phi)$ and $V_\sigma(\sigma)$ are the potentials for tachyon, quintessence and phantom fields, respectively.

We consider a spatially-flat FRW metric

$$ds^2 = dt^2 - a^2(t)[dr^2 + r^2 d\Omega^2], \tag{5}$$

and a tetrad field of the form $e^i_\mu = diag(1, a, a, a)$. Then the Friedmann equations for FRW metric read as

$$H^2 = \frac{\kappa^2}{3}(\rho_\psi + \rho_\phi + \rho_\sigma), \tag{6}$$

$$\dot{H} = -\frac{\kappa^2}{2}(\rho_\psi + p_\psi + \rho_\phi + p_\phi + \rho_\sigma + p_\sigma), \tag{7}$$

where $H = \dot{a}/a$ is the Hubble parameter, $a$ is the scale factor, dot represents the derivative with respect to cosmic time $t$. $\rho$ and $p$ are the energy density and the pressure of the corresponding scalar field constituents of the dark energy.

Conservation of energy gives the evolution equations for the dark energy constituents, such as

$$\dot{\rho}_\psi + 3H(\rho_\psi + p_\psi) = 0, \tag{8}$$

$$\dot{\rho}_\phi + 3H(\rho_\phi + p_\phi) = 0, \tag{9}$$

$$\dot{\rho}_\sigma + 3H(\rho_\sigma + p_\sigma) = 0. \tag{10}$$

The total energy density and the pressure of dark energy reads

$$\rho_{tot} = \rho_{DE} = \rho_\psi + \rho_\phi + \rho_\sigma, \tag{11}$$

$$p_{tot} = p_{DE} = p_\psi + p_\phi + p_\sigma, \tag{12}$$

with the total equation of state parameter

$$\omega_{tot} = \frac{p_{tot}}{\rho_{tot}} = \omega_{\psi}\,\Omega_{\psi} + \omega_{\phi}\,\Omega_{\phi} + \omega_{\sigma}\,\Omega_{\sigma}\,, \tag{13}$$

where $\omega_{\psi} = p_{\psi}/\rho_{\psi}$, $\omega_{\phi} = p_{\phi}/\rho_{\phi}$ and $\omega_{\sigma} = p_{\sigma}/\rho_{\sigma}$ are the equation of state parameters and $\Omega_{\psi} = \rho_{\psi}/\rho_{tot}$, $\Omega_{\phi} = \rho_{\phi}/\rho_{tot}$ and $\Omega_{\sigma} = \rho_{\sigma}/\rho_{tot}$ are the density parameters for the tachyon, quintessence and phantom fields, respectively. Then the total density parameter is defined as

$$\Omega_{tot} = \Omega_{\psi} + \Omega_{\phi} + \Omega_{\sigma} = \frac{\kappa^2 \rho_{tot}}{3H^2} = 1\,, \tag{14}$$

where we assume that three kinds of scalar fields constitute the dark energy with an equal proportion of density parameter such that, $\Omega_{\psi} = \Omega_{\phi} = \Omega_{\sigma} = 1/3$.

The Lagrangian of the scalar fields are reexpressed from the action in equation (1), as

$$L_{\psi} = \xi_1 f(\psi) T - V_{\psi}(\psi)\sqrt{1 - \frac{\dot{\psi}^2}{V_{\psi}(\psi)}}\,, \tag{15}$$

$$L_{\phi} = \xi_2 g(\phi) T + \frac{1}{2}\dot{\phi}^2 - V_{\phi}(\phi)\,, \tag{16}$$

$$L_{\sigma} = \xi_3 h(\sigma) T - \frac{1}{2}\dot{\sigma}^2 - V_{\sigma}(\sigma)\,. \tag{17}$$

Then the energy density and pressure values for three scalar fields can be found by the variation of the total Lagrangian in (1) with respect to the tetrad field $e_i^{\mu}$. After the variation of Lagrangian, there come contributions from the geometric terms so the (0,0)-component and (i,i)-component of the stress-energy tensor give the energy density and the pressure, respectively. Accordingly the energy density and pressure values for the tachyon, quintessence and phantom fields read, as

$$\rho_{\psi} = -\xi_1 f(\psi) 6H^2 + \frac{V_{\psi}(\psi)}{\sqrt{1 - (\dot{\psi}^2 / V_{\psi}(\psi))}}, \tag{18}$$

$$p_{\psi} = \xi_1 f(\psi) 6H^2 - V_{\psi}(\psi) \sqrt{1 - \frac{\dot{\psi}^2}{V_{\psi}(\psi)}} + 4\xi_1 \dot{H} f(\psi) + 4\xi_1 H f'(\psi)\dot{\psi}^2, \tag{19}$$

$$\rho_{\phi} = -\xi_2 g(\phi) 6H^2 + \frac{1}{2}\dot{\phi}^2 + V_{\phi}(\phi), \tag{20}$$

$$p_{\phi} = \xi_2 g(\phi) 6H^2 + \frac{1}{2}\dot{\phi}^2 - V_{\phi}(\phi) + 4\xi_2 \dot{H} g(\phi) + 4\xi_2 H g'(\phi)\dot{\phi}, \tag{21}$$

$$\rho_{\sigma} = -\xi_3 h(\sigma) 6H^2 - \frac{1}{2}\dot{\sigma}^2 + V_{\sigma}(\sigma), \tag{22}$$

$$p_{\sigma} = \xi_3 h(\sigma) 6H^2 - \frac{1}{2}\dot{\sigma}^2 - V_{\sigma}(\sigma) + 4\xi_3 \dot{H} h(\sigma) + 4\xi_3 H h'(\sigma)\dot{\sigma}, \tag{23}$$

respectively. Here we have used the relation $T = -6H^2$, and prime denotes the derivative of related coupling functions with respect to the related field variables. Now, we can find the equation of motions for three scalar fields from the variation of the field Lagrangians (15)-(17) with respect to the field variables $\psi$, $\phi$ and $\sigma$, such that

$$\ddot{\psi} + 3H\dot{\psi}\left(1 - \frac{\dot{\psi}^2}{V_{\psi}(\psi)}\right) + \left(1 - \frac{3}{2}\frac{\dot{\psi}^2}{V_{\psi}(\psi)}\right)V_{\psi}(\psi) + 6\xi_1\left(1 - \frac{\dot{\psi}^2}{V_{\psi}(\psi)}\right)^{\frac{3}{2}}H^2 f'(\psi) = 0, \tag{24}$$

$$\ddot{\phi} + V'_{\phi}(\phi) + 6\xi_2 H^2 g'(\phi) + 3H\dot{\phi} = 0, \tag{25}$$

$$\ddot{\sigma} - V'_{\sigma}(\sigma) - 6\xi_3 H^2 h'(\sigma) + 3H\dot{\sigma} = 0. \tag{26}$$

These equations of motions are for the tachyon, quintessence and phantom constituents of dark energy, respectively. Here, the prime of potentials denotes the derivative of related field potentials with respect to the related field variables. All these evolution equations in (24)-(26) can also be obtained by using the relations (18)-(23) in the continuity equations (8)-(10).

We now perform the phase-space analysis of the model in order to investigate the late-time solutions of the universe considered here.

## 3. Phase-space and stability analysis

We study the properties of the constructed dark energy model by performing the phase-space analysis. Therefore we transform the aforementioned dynamical system into its autonomous form [38,39,43-46]. To proceed we introduce the auxiliary variables

$$x_\psi = \frac{\dot{\psi}}{\sqrt{V_\psi(\psi)}}, \qquad y_\psi = \frac{\kappa\sqrt{V_\psi(\psi)}}{\sqrt{3}H}, \qquad u_\psi = \kappa\sqrt{f(\psi)}, \tag{27}$$

$$x_\phi = \frac{\kappa\dot{\phi}}{\sqrt{6}H}, \qquad y_\phi = \frac{\kappa\sqrt{V_\phi(\phi)}}{\sqrt{3}H}, \qquad u_\phi = \kappa\sqrt{g(\phi)}, \tag{28}$$

$$x_\sigma = \frac{\kappa\dot{\sigma}}{\sqrt{6}H}, \qquad y_\sigma = \frac{\kappa\sqrt{V_\sigma(\sigma)}}{\sqrt{3}H}, \qquad u_\sigma = \kappa\sqrt{h(\sigma)}, \tag{29}$$

together with $N = \ln a$ and for any quantity $F$, the time derivative is $\dot{F} = H(dF/dN)$.

We rewrite the density parameters for the fields $\psi$, $\phi$ and $\sigma$ in the autonomous system by using (14), (18), (20) and (22) with (27)-(29)

$$\Omega_\psi = \frac{\kappa^2\rho_\psi}{3H^2} = \mu\, y_\psi^2 - 2\xi_1 u_\psi^2, \tag{30}$$

$$\Omega_\phi = \frac{\kappa^2\rho_\phi}{3H^2} = x_\phi^2 + y_\phi^2 - 2\xi_2 u_\phi^2, \tag{31}$$

$$\Omega_\sigma = \frac{\kappa^2\rho_\sigma}{3H^2} = -x_\sigma^2 + y_\sigma^2 - 2\xi_3 u_\sigma^2, \tag{32}$$

and the total density parameter is

$$\Omega_{tot} = \frac{\kappa^2 \rho_{tot}}{3H^2} = \Omega_\psi + \Omega_\phi + \Omega_\sigma$$
$$= \mu\, y_\psi^2 - 2\xi_1 u_\psi^2 + x_\phi^2 + y_\phi^2 - 2\xi_2 u_\phi^2 - x_\sigma^2 + y_\sigma^2 - 2\xi_3 u_\sigma^2 = 1 \qquad (33)$$

where $\mu = 1/\sqrt{1 - (\dot{\psi}^2/V_\psi(\psi))} = 1/\sqrt{1 - x_\psi^2}$. Then the equation of state parameters can be written in the autonomous form by using (18)-(23) in $\omega = p/\rho$ for every scalar field, such as

$$\omega_\psi = \frac{p_\psi}{\rho_\psi} = \frac{-\mu^{-1} y_\psi^2 + 2\xi_1 u_\psi \left[ \frac{2\sqrt{3}}{3} \alpha_\psi x_\psi y_\psi + u_\psi (1 - \frac{2}{3}s) \right]}{\mu\, y_\psi^2 - 2\xi_1 u_\psi^2}, \qquad (34)$$

$$\omega_\phi = \frac{p_\phi}{\rho_\phi} = \frac{x_\phi^2 - y_\phi^2 + 2\xi_2 u_\phi \left[ \frac{2\sqrt{6}}{3} \alpha_\phi x_\phi + u_\phi (1 - \frac{2}{3}s) \right]}{x_\phi^2 + y_\phi^2 - 2\xi_2 u_\phi^2}, \qquad (35)$$

$$\omega_\sigma = \frac{p_\sigma}{\rho_\sigma} = \frac{-x_\sigma^2 - y_\sigma^2 + 2\xi_3 u_\sigma \left[ \frac{2\sqrt{6}}{3} \alpha_\sigma x_\sigma + u_\sigma (1 - \frac{2}{3}s) \right]}{-x_\sigma^2 + y_\sigma^2 - 2\xi_3 u_\sigma^2}, \qquad (36)$$

where $\alpha_\psi = f'(\psi)/\sqrt{f(\psi)}$, $\alpha_\phi = g'(\phi)/\sqrt{g(\phi)}$, $\alpha_\sigma = h'(\sigma)/\sqrt{h(\sigma)}$ and $s = -\dot{H}/H^2$. From (13) and (30)-(32) and (34)-(36), we obtain $\omega_{tot}$ in the autonomous system, as

$$\omega_{tot} = -\mu^{-1} y_\psi^2 + 2\xi_1 u_\psi \left[ \frac{2\sqrt{3}}{3} \alpha_\psi x_\psi y_\psi + u_\psi (1 - \frac{2}{3}s) \right]$$
$$+ x_\phi^2 - y_\phi^2 + 2\xi_2 u_\phi \left[ \frac{2\sqrt{6}}{3} \alpha_\phi x_\phi + u_\phi (1 - \frac{2}{3}s) \right] \qquad . \qquad (37)$$
$$- x_\sigma^2 - y_\sigma^2 + 2\xi_3 u_\sigma \left[ \frac{2\sqrt{6}}{3} \alpha_\sigma x_\sigma + u_\sigma (1 - \frac{2}{3}s) \right]$$

We can express $s$ in the autonomous system by using (6), (7) and (37), such that

$$s = -\frac{\dot{H}}{H^2} = \frac{3}{2}(1+\omega_{tot}) = \frac{3}{2} - \frac{3}{2}\mu^{-1}y_\psi^2 + \xi_1 u_\psi \left[2\sqrt{3}\alpha_\psi x_\psi y_\psi + u_\psi(3-2s)\right]$$
$$+ \frac{3}{2}x_\phi^2 - \frac{3}{2}y_\phi^2 + \xi_2 u_\phi \left[2\sqrt{6}\alpha_\phi x_\phi + u_\phi(3-2s)\right] \qquad (38)$$
$$- \frac{3}{2}x_\sigma^2 - \frac{3}{2}y_\sigma^2 + \xi_3 u_\sigma \left[2\sqrt{6}\alpha_\sigma x_\sigma + u_\sigma(3-2s)\right]$$

and

$$s = \left\{ \frac{3}{2} - \frac{3}{2}\mu^{-1}y_\psi^2 + \xi_1 u_\psi \left[2\sqrt{3}\alpha_\psi x_\psi y_\psi + 3u_\psi\right] \right.$$
$$+ \frac{3}{2}x_\phi^2 - \frac{3}{2}y_\phi^2 + \xi_2 u_\phi \left[2\sqrt{6}\alpha_\phi x_\phi + 3u_\phi\right] \qquad . \qquad (39)$$
$$\left. - \frac{3}{2}x_\sigma^2 - \frac{3}{2}y_\sigma^2 + \xi_3 u_\sigma \left[2\sqrt{6}\alpha_\sigma x_\sigma + 3u_\sigma\right] \right\} \left[1 + 2\xi_1 u_\psi^2 + 2\xi_2 u_\phi^2 + 2\xi_3 u_\sigma^2\right]^{-1}$$

Here $s$ is only a jerk parameter used in other equations of cosmological parameters. Then the deceleration parameter $q$ is

$$q = -1 - \frac{\dot{H}}{H^2} = \frac{1}{2} + \frac{3}{2}\omega_{tot} = \frac{1}{2} - \frac{3}{2}\mu^{-1}y_\psi^2 + \xi_1 u_\psi \left[2\sqrt{3}\alpha_\psi x_\psi y_\psi + u_\psi(3-2s)\right]$$
$$+ \frac{3}{2}x_\phi^2 - \frac{3}{2}y_\phi^2 + \xi_2 u_\phi \left[2\sqrt{6}\alpha_\phi x_\phi + u_\phi(3-2s)\right] \qquad . \qquad (40)$$
$$- \frac{3}{2}x_\sigma^2 - \frac{3}{2}y_\sigma^2 + \xi_3 u_\sigma \left[2\sqrt{6}\alpha_\sigma x_\sigma + u_\sigma(3-2s)\right]$$

Now we transform the equations of motions (6), (7) and (24)-(26) into the autonomous system containing the auxiliary variables in (27)-(29) and their derivatives with respect to $N = \ln a$. Thus we obtain $X' = f(X)$, where $X$ is the column vector including the auxiliary variables and $f(X)$ is the column vector of the autonomous equations. After writing $X'$, we find the critical points $X_c$ of $X$, by setting $X' = 0$. We expand $X' = f(X)$ around $X = X_c + U$, where $U$ is the column vector of perturbations of the auxiliary variables, such as $\delta x$, $\delta y$ and $\delta u$ for each scalar field. Thus, we expand the perturbation equations up to the first order for each critical point as $U' = MU$, where $M$ is the matrix of perturbation equations. For each critical points, the eigenvalues of perturbation matrix $M$ determine the type and stability of the critical points [47-50].

Particularly, the autonomous form of the cosmological system in (6), (7) and (24)-(26) is [51-60]:

$$x'_\psi = \frac{\sqrt{3}}{2}\Big[\lambda_\psi x_\psi^2 y_\psi + \lambda_\psi(2-3x_\psi^2)y_\psi - 4\alpha_\psi \xi_1 u_\psi \mu^{-3} y_\psi^{-1} - 2\sqrt{3}x_\psi \mu^{-2}\Big], \tag{41}$$

$$y'_\psi = \Big[-\frac{\sqrt{3}}{2}\lambda_\psi x_\psi y_\psi + s\Big]y_\psi, \tag{42}$$

$$u'_\psi = \frac{\sqrt{3}}{2}\alpha_\psi x_\psi y_\psi, \tag{43}$$

$$\begin{aligned}
x'_\phi = -3x_\phi\Big[&1+x_\sigma^2-x_\phi^2+\frac{2}{3}\xi_2(su_\phi-\sqrt{6}\alpha_\phi x_\phi)u_\phi+\frac{2}{3}\xi_3(su_\sigma-\sqrt{6}\alpha_\sigma x_\sigma)u_\sigma\\
&-\frac{\mu}{2}x_\psi^2 y_\psi^2+\frac{2}{3}\xi_1(su_\psi-\sqrt{6}\,\alpha_\psi x_\psi)u_\psi\Big]+\lambda_\phi\frac{\sqrt{6}}{2}y_\phi^2-\sqrt{6}\xi_2\alpha_\phi u_\phi
\end{aligned}, \tag{44}$$

$$\begin{aligned}
y'_\phi = 3y_\phi\Big[&\frac{\mu}{2}x_\psi^2 y_\psi^2-\frac{2}{3}\xi_1(su_\psi-\sqrt{6}\,\alpha_\psi x_\psi)u_\psi+x_\phi^2-\frac{2}{3}\xi_2(su_\phi-\sqrt{6}\alpha_\phi x_\phi)u_\phi\\
&-x_\sigma^2-\frac{2}{3}\xi_3(su_\sigma-\sqrt{6}\alpha_\sigma x_\sigma)u_\sigma-\lambda_\phi\frac{\sqrt{6}}{6}x_\phi^2\Big]
\end{aligned}, \tag{45}$$

$$u'_\phi = \frac{\sqrt{6}}{2}\alpha_\phi x_\phi, \tag{46}$$

$$\begin{aligned}
x'_\sigma = -3x_\sigma\Big[&1+x_\sigma^2-x_\phi^2+\frac{2}{3}\xi_2(su_\phi-\sqrt{6}\alpha_\phi x_\phi)u_\phi+\frac{2}{3}\xi_3(su_\sigma-\sqrt{6}\alpha_\sigma x_\sigma)u_\sigma\\
&-\frac{\mu}{2}x_\psi^2 y_\psi^2+\frac{2}{3}\xi_1(su_\psi-\sqrt{6}\,\alpha_\psi x_\psi)u_\psi\Big]+\lambda_\sigma\frac{\sqrt{6}}{2}y_\sigma^2+\sqrt{6}\xi_3\alpha_\sigma u_\sigma
\end{aligned}, \tag{47}$$

$$y'_\sigma = 3y_\sigma \left[ \frac{\mu}{2} x_\psi^2 y_\psi^2 - \frac{2}{3}\xi_1(su_\psi - \sqrt{6}\,\alpha_\psi x_\psi)u_\psi + x_\phi^2 - \frac{2}{3}\xi_2(su_\phi - \sqrt{6}\alpha_\phi x_\phi)u_\phi \right. $$
$$\left. - x_\sigma^2 - \frac{2}{3}\xi_3(su_\sigma - \sqrt{6}\alpha_\sigma x_\sigma)u_\sigma - \lambda_\sigma \frac{\sqrt{6}}{6} x_\sigma^2 \right], \qquad (48)$$

$$u'_\sigma = \frac{\sqrt{6}}{2}\alpha_\sigma x_\sigma , \qquad (49)$$

where $\lambda_\psi = -V'_\psi / \kappa V_\psi$, $\lambda_\phi = -V'_\phi / \kappa V_\phi$ and $\lambda_\sigma = -V'_\sigma / \kappa V_\sigma$. Henceforth, we assume the non-minimal coupling functions $f(\psi) \propto \psi^2$, $g(\phi) \propto \phi^2$ and $h(\sigma) \propto \sigma^2$, thus $\alpha_\psi$, $\alpha_\phi$ and $\alpha_\sigma$ are constant. Also the usual assumption in the literature is to take the potentials $V_\psi = V_{\psi 0} e^{-k_\psi \lambda_\psi \psi}$, $V_\phi = V_{\phi 0} e^{-k_\phi \lambda_\phi \phi}$ and $V_\sigma = V_{\sigma 0} e^{-k_\sigma \lambda_\sigma \sigma}$ [61-64]. Such potentials give also constant $\lambda_\psi$, $\lambda_\phi$ and $\lambda_\sigma$.

Now we perform the phase-space analysis of the model by finding the critical points of the autonomous system in (41)-(49). To obtain these points, we set the left hand sides of the equations (41)-(49) to zero. After some calculations, four critical points are found by assuming $\omega_{tot}$ and $q$ as -1 for each critical point. The critical points are listed in Table 1 with the existence conditions.

Then we insert linear perturbations $x \to x_c + \delta x$, $y \to y_c + \delta y$ and $u \to u_c + \delta u$ about the critical points for three scalar fields $\psi$, $\phi$ and $\sigma$ in the autonomous system (41)-(49). Thus we obtain a $9\times9$ perturbation matrix $M$ whose elements are given, as

$$M_{11} = -3, \qquad (50)$$

$$M_{12} = \sqrt{3}\lambda_\psi + 2\sqrt{3}\alpha_\psi \xi_1 y_\psi^{-2} u_\psi , \qquad (51)$$

$$M_{13} = -2\sqrt{3}\alpha_\psi \xi_1 y_\psi^{-1} , \qquad (52)$$

$$M_{21} = \frac{2\sqrt{3}\alpha_\psi \xi_1}{P} u_\psi\, y_\psi^2 - \frac{\sqrt{3}}{2}\lambda_\psi\, y_\psi^2\,, \tag{53}$$

$$M_{22} = -\frac{3}{P} y_\psi^2\,, \tag{54}$$

$$M_{23} = \frac{6\xi_1}{P} u_\psi\, y_\psi\,, \tag{55}$$

$$M_{24} = \frac{2\sqrt{6}\alpha_\psi \xi_2}{P} u_\phi\, y_\psi\,, \tag{56}$$

$$M_{25} = -\frac{3}{P} y_\phi\, y_\psi\,, \tag{57}$$

$$M_{26} = \frac{6\xi_2}{P} u_\phi\, y_\psi\,, \tag{58}$$

$$M_{27} = \frac{2\sqrt{6}\,\alpha_\sigma \xi_3}{P} u_\sigma\, y_\psi\,, \tag{59}$$

$$M_{28} = -\frac{3}{P} y_\sigma\, y_\psi\,, \tag{60}$$

$$M_{29} = \frac{6\xi_3}{P} u_\sigma\, y_\psi\,, \tag{61}$$

$$M_{31} = \frac{\sqrt{3}}{2}\alpha_\psi\, y_\psi\,, \tag{62}$$

$$M_{44} = -3\,, \tag{63}$$

$$M_{45} = \sqrt{6}\lambda_\phi\, y_\phi\,, \tag{64}$$

$$M_{46} = -\sqrt{6}\,\alpha_\phi \xi_2 \,, \tag{65}$$

$$M_{51} = 2\sqrt{6}\,\alpha_\psi \xi_1 u_\psi\, y_\phi - \frac{4\sqrt{3}\alpha_\psi \xi_1}{P}\left(\frac{P-1}{2}\right)u_\psi\, y_\psi\, y_\phi \,, \tag{66}$$

$$M_{52} = \frac{6}{P}\left(\frac{P-1}{2}\right)y_\psi\, y_\phi \,, \tag{67}$$

$$M_{53} = -\frac{12\xi_1}{P}\left(\frac{P-1}{2}\right)u_\psi\, y_\phi \,, \tag{68}$$

$$M_{54} = 2\sqrt{6}\,\alpha_\phi \xi_2 u_\phi\, y_\phi - \frac{4\sqrt{6}\alpha_\phi \xi_2}{P}\left(\frac{P-1}{2}\right)u_\phi\, y_\phi - \frac{\sqrt{6}\lambda_\phi}{2}\, y_\phi \,, \tag{69}$$

$$M_{55} = \frac{6}{P}\left(\frac{P-1}{2}\right)y_\phi^2 \,, \tag{70}$$

$$M_{56} = -\frac{12\xi_2}{P}\left(\frac{P-1}{2}\right)u_\phi\, y_\phi \,, \tag{71}$$

$$M_{57} = 2\sqrt{6}\,\alpha_\sigma \xi_3 u_\sigma\, y_\phi - \frac{4\sqrt{6}\alpha_\sigma \xi_3}{P}\left(\frac{P-1}{2}\right)u_\sigma\, y_\phi \,, \tag{72}$$

$$M_{58} = \frac{6}{P}\left(\frac{P-1}{2}\right)y_\sigma\, y_\phi \,, \tag{73}$$

$$M_{59} = -\frac{12\xi_3}{P}\left(\frac{P-1}{2}\right)u_\sigma\, y_\phi \,, \tag{74}$$

$$M_{64} = \frac{\sqrt{6}}{2}\,\alpha_\phi \,, \tag{75}$$

$$M_{77} = -3, \tag{76}$$

$$M_{78} = -\sqrt{6}\lambda_\sigma y_\sigma, \tag{77}$$

$$M_{79} = \sqrt{6}\alpha_\sigma \xi_3, \tag{78}$$

$$M_{81} = 2\sqrt{6}\,\alpha_\psi \xi_1 u_\psi y_\sigma - \frac{4\sqrt{3}\alpha_\psi \xi_1}{P}\left(\frac{P-1}{2}\right) u_\psi y_\psi y_\sigma, \tag{79}$$

$$M_{82} = \frac{6}{P}\left(\frac{P-1}{2}\right) y_\psi y_\sigma, \tag{80}$$

$$M_{83} = -\frac{12\xi_1}{P}\left(\frac{P-1}{2}\right) u_\psi y_\sigma, \tag{81}$$

$$M_{84} = 2\sqrt{6}\,\alpha_\phi \xi_2 u_\phi y_\sigma - \frac{4\sqrt{6}\alpha_\phi \xi_2}{P}\left(\frac{P-1}{2}\right) u_\phi y_\sigma, \tag{82}$$

$$M_{85} = \frac{6}{P}\left(\frac{P-1}{2}\right) y_\phi y_\sigma, \tag{83}$$

$$M_{86} = -\frac{12\xi_2}{P}\left(\frac{P-1}{2}\right) u_\phi y_\sigma, \tag{84}$$

$$M_{87} = 2\sqrt{6}\,\alpha_\sigma \xi_3 u_\sigma y_\sigma - \frac{4\sqrt{6}\alpha_\sigma \xi_3}{P}\left(\frac{P-1}{2}\right) u_\sigma y_\sigma - \frac{\sqrt{6}\lambda_\sigma}{2} y_\sigma, \tag{85}$$

$$M_{88} = \frac{6}{P}\left(\frac{P-1}{2}\right) y_\sigma^2, \tag{86}$$

$$M_{89} = -\frac{12\xi_3}{P}\left(\frac{P-1}{2}\right)u_\sigma y_\sigma, \tag{87}$$

$$M_{97} = \frac{\sqrt{6}}{2}\alpha_\sigma, \tag{88}$$

and all the other matrix elements except given above in (50)-(88) are zero. Here also $P = 1 + 2\xi_1 u_\psi^2 + 2\xi_2 u_\phi^2 + 2\xi_3 u_\sigma^2$. Then we calculate the eigenvalues of perturbation matrix $M$ for four critical points given in Table 1 with the corresponding existing conditions. We obtain four sets of eigenvalues for four perturbation matrices for each of four critical points. To determine the type and stability of critical points, we examine the sign of the real parts of eigenvalues. The critical point is stable for negative real part of eigenvalues. The physical meaning of negative eigenvalue is always a late-time stable attractor, namely if the universe reaches, it keeps its state forever and thus it can attract the universe at a late-time. Accelerated expansion occurs here, because $\omega_{tot} = -1 < -1/3$. If a convenient condition is provided, an accelerated contraction can even exist for $\omega_{tot} = -1 < -1/3$ value. Eigenvalues of the matrix $M$ are represented in Table 2 for each of the critical points $A$, $B$, $C$ and $D$.

As seen in Table 2, the first two critical points $A$ and $B$ have same eigenvalues as $C$ and $D$ have same eigenvalues. The stability conditions of each critical point are listed in Table 3, according to the sign of the eigenvalues.

In order to analyze the cosmological behavior of each critical point, the attractor solutions in scalar field cosmology should be noted [65]. In modern theoretical cosmology it is common that the energy density of one or more scalar fields exerts a crucial influence on the evolution of the universe and some certain conditions or behaviors naturally affect this evolution. The evolution and the affecting factors on this evolution meet in the term of cosmological attractors: the scalar field evolution approaches a certain kind of behavior by the dynamical conditions without finely tuned initial conditions [66–78], either in inflationary cosmology or late-time dark energy models. Attractor behavior is a situation in which a collection of phase-space points evolve into a certain region and never leave.

*Critical point A:* This point exists for all values of $\xi_1$, $\xi_2$, $\xi_3$, $\alpha_\psi$, $\alpha_\phi$ and $\alpha_\sigma$ while $\lambda_\psi = \lambda_\phi = \lambda_\sigma = 0$ which means the potentials $V_\psi$, $V_\phi$ and $V_\sigma$ are constant. Acceleration occurs at this point since $\omega_{tot} = -1 < -1/3$ and this is an expansion phase because $y_\psi$, $y_\phi$ and $y_\sigma$ have positive values, so $H$ does. Point $A$ is stable (meaning that universe keeps its evolution), if $0 < 3/(4\alpha_\psi^2) < \xi_1$, $0 < 3/(4\alpha_\phi^2) < \xi_2$ and $\xi_3 < -3/(4\alpha_\sigma^2) < 0$, but it is a saddle point (meaning that the universe evolves between different states) for values $\xi_1, \xi_2 < 0$. In Figure 1, we represent the 2-dimensional projections of 9-dimensional phase-space trajectories for the values of $\xi_1 = \xi_2 = -\xi_3 = 0.5$, $\alpha_\psi = \alpha_\phi = -\alpha_\sigma = 1.5$ and three auxiliary $\lambda$ values for each field $\psi$, $\phi$ and $\sigma$. This state corresponds to a stable attractor starting from the critical point $A$, as seen from the plots in Figure 1.

*Critical point B:* Point $B$ also exists for any values of $\xi_1$, $\xi_2$, $\xi_3$, $\alpha_\psi$, $\alpha_\phi$ and $\alpha_\sigma$ while $\lambda_\psi = \lambda_\phi = \lambda_\sigma = 0$ which means the potentials $V_\psi$, $V_\phi$ and $V_\sigma$ are constant. Acceleration phase is again valid here since $\omega_{tot} = -1 < -1/3$, but this point refers to contraction phase because $y_\psi$, $y_\phi$ and $y_\sigma$ have negative sign. Stability of the point $B$ is same with the point $A$ for same conditions. Therefore the stable attractor behavior for contraction is represented in Figure 2. We plot 2-dimensional projections of phase-space trajectories for same values as point $A$, $\xi_1 = \xi_2 = -\xi_3 = 0.5$, $\alpha_\psi = \alpha_\phi = -\alpha_\sigma = 1.5$ and auxiliary $\lambda$ values.

*Critical point C:* Critical point $C$ occurs for $\xi_1, \xi_2, \xi_3 < 0$ and $\lambda_\psi, \lambda_\phi, \lambda_\sigma < 0$ while $\alpha_\psi = \alpha_\phi = \alpha_\sigma = 0$ meaning constant coupling functions $f(\psi)$, $g(\phi)$ and $h(\sigma)$. The cosmological behavior is again an acceleration phase since $\omega_{tot} < -1/3$. Point $C$ is stable for any values of $\xi$, $\alpha$ and $\lambda$ of fields $\psi$, $\phi$ and $\sigma$. 2-dimensional projections of phase-space are represented in Figure 3 for $\xi_1 = \xi_2 = \xi_3 = -0.5$, $\alpha_\psi = \alpha_\phi = \alpha_\sigma = -1.5$ and three auxiliary $\lambda$ values for each field $\psi$, $\phi$ and $\sigma$. A stable attractor starting from the critical point $C$ is seen from the plots in Figure 3.

*Critical point D:* This point exists for $\xi_1, \xi_2, \xi_3 < 0$ and $\lambda_\psi, \lambda_\phi, \lambda_\sigma > 0$ while $\alpha_\psi = \alpha_\phi = \alpha_\sigma = 0$ implying constant $f(\psi)$, $g(\phi)$, $h(\sigma)$ coupling functions. Acceleration phase is valid due to $\omega_{tot} < -1/3$. Point $D$ is also stable for all $\xi$, $\alpha$ and $\lambda$ values of fields $\psi$, $\phi$ and $\sigma$. 2-dimensional plots of phase-space trajectories are shown in Figure 4 for $\xi_1 = \xi_2 = \xi_3 = -0.5$, $\alpha_\psi = \alpha_\phi = \alpha_\sigma = -1.5$ and three auxiliary $\lambda$ values for each field $\psi$, $\phi$ and $\sigma$. This state again corresponds to a stable attractor starting from the point $C$, as in Figure 4.

All the plots in Figures 1-4 has the structure of stable attractor, since each of them evolves to a single point which is in fact one of the critical points in Table 1. These evolutions to the critical points are the attractor solutions in mixed-dark energy domination cosmology of our model which imply an expanding universe.

## 4. Conclusion

Mixed dark energy is a generalized combination of tachyon, quintessence and phantom fields non-minimally coupled to gravity [33,34,38,39]. These three scalar fields are the constituents of mixed dark energy. Firstly, the action integral of non-minimally coupled mixed dark energy model is set up to study its dynamics. Here we consider that our dark energy constituents interact only with the gravity. There could also be chosen some interactions between the dark constituents, but in order to start from a pedagogical order we prefer to exclude the dark interactions. We obtain the Hubble parameter and Friedmann equations of model in spatially-flat FRW geometry. Energy density and pressure values with the evolution equations for tachyon, quintessence and phantom fields are obtained from the variation of action and Lagrangian of model. Then we translate these dynamical expressions into the autonomous form by introducing suitable auxiliary variables in order to perform the phase-space analysis. We find the critical points of autonomous system by setting each autonomous equation to zero. By constructing the perturbation equations, we find four perturbation matrices for each critical point. The eigenvalues of four perturbation matrices are determined to examine the stability of critical points. We also calculate some important cosmological parameters, for instance the total equation of state parameter and deceleration parameter to see whether the critical points correspond to an accelerating universe or not. There are four stable attractors of model depending on the non-minimal coupling parameters $\xi_1$, $\xi_2$, $\xi_3$ and $\alpha_\psi$, $\alpha_\phi$, $\alpha_\sigma$ values. All of the stable solutions correspond to an accelerating universe due to $\omega_{tot} < -1/3$. For constant potentials $V_\psi$, $V_\phi$ and $V_\sigma$ the critical points $A$ and $B$ are late-time stable attractors for $0 < 3/(4\alpha_\psi^2) < \xi_1$, $0 < 3/(4\alpha_\phi^2) < \xi_2$ and $\xi_3 < -3/(4\alpha_\sigma^2) < 0$. While point $A$ refers to an expansion with a stable acceleration, point $B$ refers to a contraction. However, for constant coupling functions $f(\psi)$, $g(\phi)$, $h(\sigma)$ the critical points $C$ and $D$ are stable attractors for any values of $\xi_1$, $\xi_2$, $\xi_3$, $\alpha_\psi$, $\alpha_\phi$ and $\alpha_\sigma$. The behavior of the model at each critical point being a stable attractor is demonstrated in Figures 1-4. In order to plot the graphs in Figures 1-4, we use adaptive Runge-Kutta method of 4th and 5th order to solve differential equations (41-49) in Matlab. Solutions for the equations with stability conditions of critical points are plotted for each pair of the solution being the auxiliary variables in (27-29).

These figures show that by choosing parameters of the model depending on the existence conditions of critical points $A$, $B$, $C$ and $D$, we obtain the attractors of the model as $A$, $B$, $C$ and $D$. Then depending on the stability conditions, suitable parameters give stable behavior for each attractor of the model. The results are consistent with the observed and expected behavior of the universe in which some epochs correspond to an accelerating expansion phase, and some will correspond to an accelerating contraction in future times [1-9, 77,78].

**Conflict of Interests**

The authors declare that there is no conflict of interests regarding the publication of this paper.

TABLE 1: Critical points and existence conditions

| Label | $x_\psi$ | $y_\psi$ | $u_\psi$ | $x_\phi$ | $y_\phi$ | $u_\phi$ | $x_\sigma$ | $y_\sigma$ | $u_\sigma$ | $\omega_{tot}$ | $q$ | Existence |
|---|---|---|---|---|---|---|---|---|---|---|---|---|
| $A$ | 0 | $\dfrac{1}{\sqrt{3}}$ | 0 | 0 | $\dfrac{1}{\sqrt{3}}$ | 0 | 0 | $\dfrac{1}{\sqrt{3}}$ | 0 | $-1$ | $-1$ | $\lambda_\psi = \lambda_\phi = \lambda_\sigma = 0$ |
| $B$ | 0 | $\dfrac{-1}{\sqrt{3}}$ | 0 | 0 | $\dfrac{-1}{\sqrt{3}}$ | 0 | 0 | $\dfrac{-1}{\sqrt{3}}$ | 0 | $-1$ | $-1$ | $\lambda_\psi = \lambda_\phi = \lambda_\sigma = 0$ |
| $C$ | 0 | 0 | $\dfrac{1}{\sqrt{6|\xi_1|}}$ | 0 | 0 | $\dfrac{1}{\sqrt{6|\xi_2|}}$ | 0 | 0 | $\dfrac{1}{\sqrt{6|\xi_3|}}$ | $-1$ | $-1$ | $\alpha_\psi = \alpha_\phi = \alpha_\sigma = 0$ and $\xi_1, \xi_2, \xi_3 < 0$, $\lambda_\psi, \lambda_\phi, \lambda_\sigma < 0$ |
| $D$ | 0 | 0 | $\dfrac{-1}{\sqrt{6|\xi_1|}}$ | 0 | 0 | $\dfrac{-1}{\sqrt{6|\xi_2|}}$ | 0 | 0 | $\dfrac{-1}{\sqrt{6|\xi_3|}}$ | $-1$ | $-1$ | $\alpha_\psi = \alpha_\phi = \alpha_\sigma = 0$ and $\xi_1, \xi_2, \xi_3 < 0$, $\lambda_\psi, \lambda_\phi, \lambda_\sigma > 0$ |

TABLE 2: Eigenvalues of the perturbation matrix for each critical point

| Critical Points | Eigenvalues |
|---|---|
| $A$ and $B$ | $\frac{1}{2}\sqrt{9-12\xi_1\alpha_\psi^2}-\frac{3}{2},$ $-\frac{1}{2}\sqrt{9-12\xi_1\alpha_\psi^2}-\frac{3}{2},$ $\frac{1}{2}\sqrt{9-12\xi_2\alpha_\phi^2}-\frac{3}{2},$ $-\frac{1}{2}\sqrt{9-12\xi_2\alpha_\phi^2}-\frac{3}{2},$ $\frac{1}{2}\sqrt{12\xi_3\alpha_\sigma^2+9}-\frac{3}{2},$ $-\frac{1}{2}\sqrt{12\xi_3\alpha_\sigma^2+9}-\frac{3}{2},$ $-1$ |
| $C$ and $D$ | $-3$ |

TABLE 3: Stability of the critical points

| Critical Points | Stability |
|---|---|
| $A$ and $B$ | Stable point, if $0 < \dfrac{3}{4\alpha_\psi^2} < \xi_1$, $0 < \dfrac{3}{4\alpha_\phi^2} < \xi_2$<br><br>and $\xi_3 < \dfrac{-3}{4\alpha_\sigma^2} < 0$<br><br><br><br>Saddle point, if $\xi_1, \xi_2 < 0$ |
| $C$ and $D$ | Stable point for all $\xi$, $\alpha$ and $\lambda$ |

FIGURE 1: Two dimensional projections of the phase-space trajectories for $\xi_1 = \xi_2 = -\xi_3 = 0.5$, $\alpha_\psi = \alpha_\phi = -\alpha_\sigma = 1.5$. All plots begin from the critical point $A$ being a stable attractor.

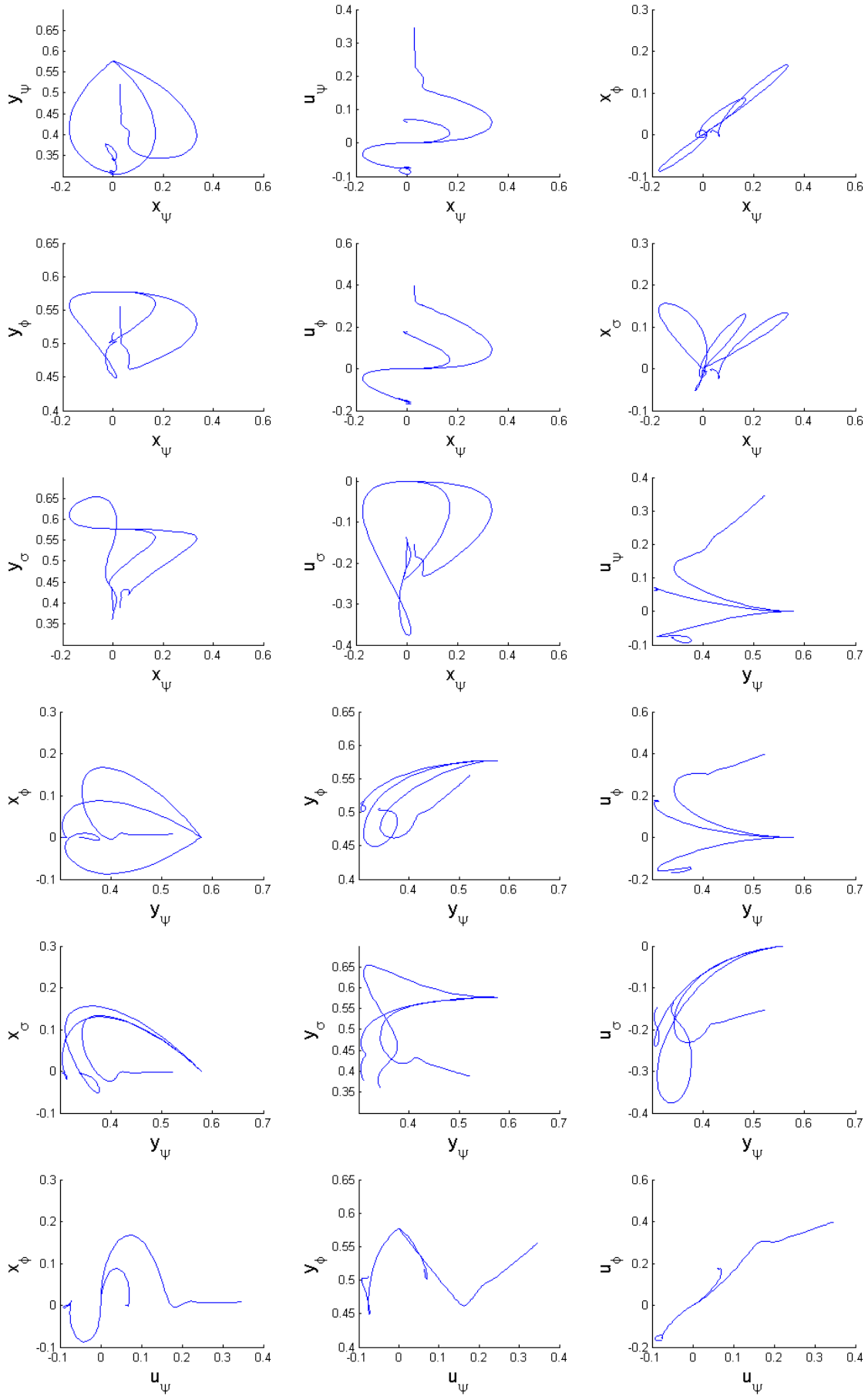

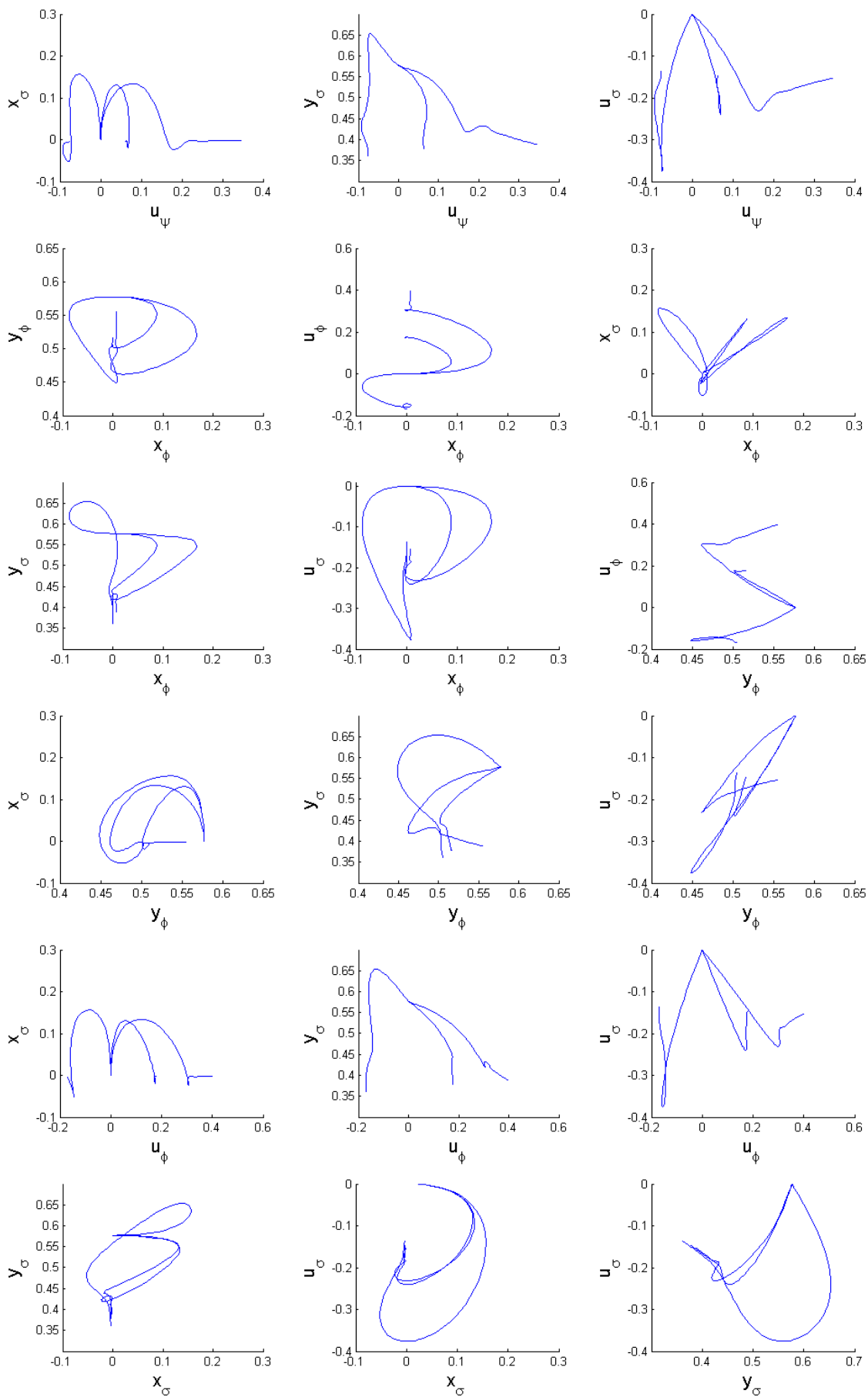

FIGURE 2: Two dimensional projections of the phase-space trajectories for $\xi_1 = \xi_2 = -\xi_3 = 0.5$, $\alpha_\psi = \alpha_\phi = -\alpha_\sigma = 1.5$. All plots begin from the critical point $B$ being a stable attractor.

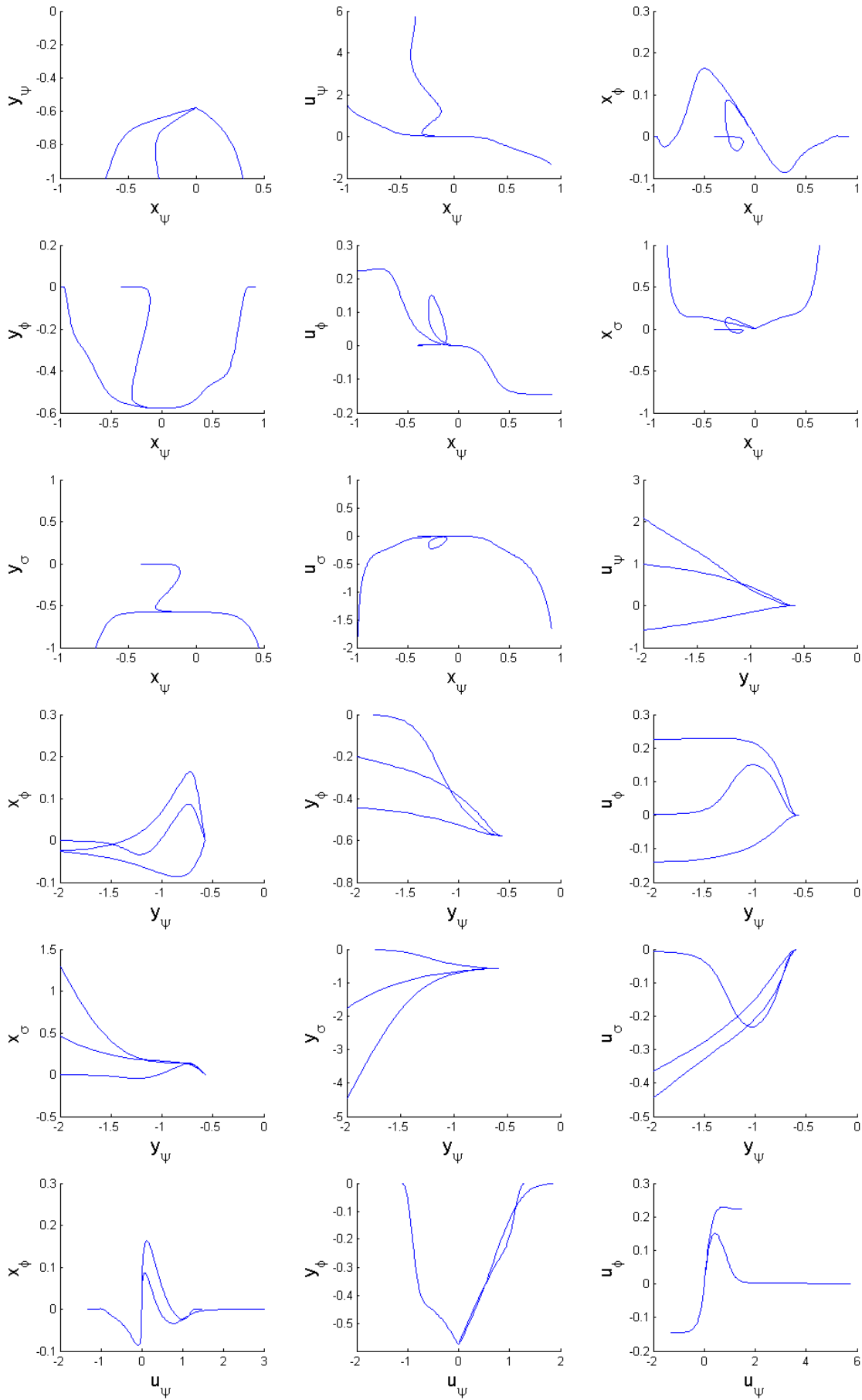

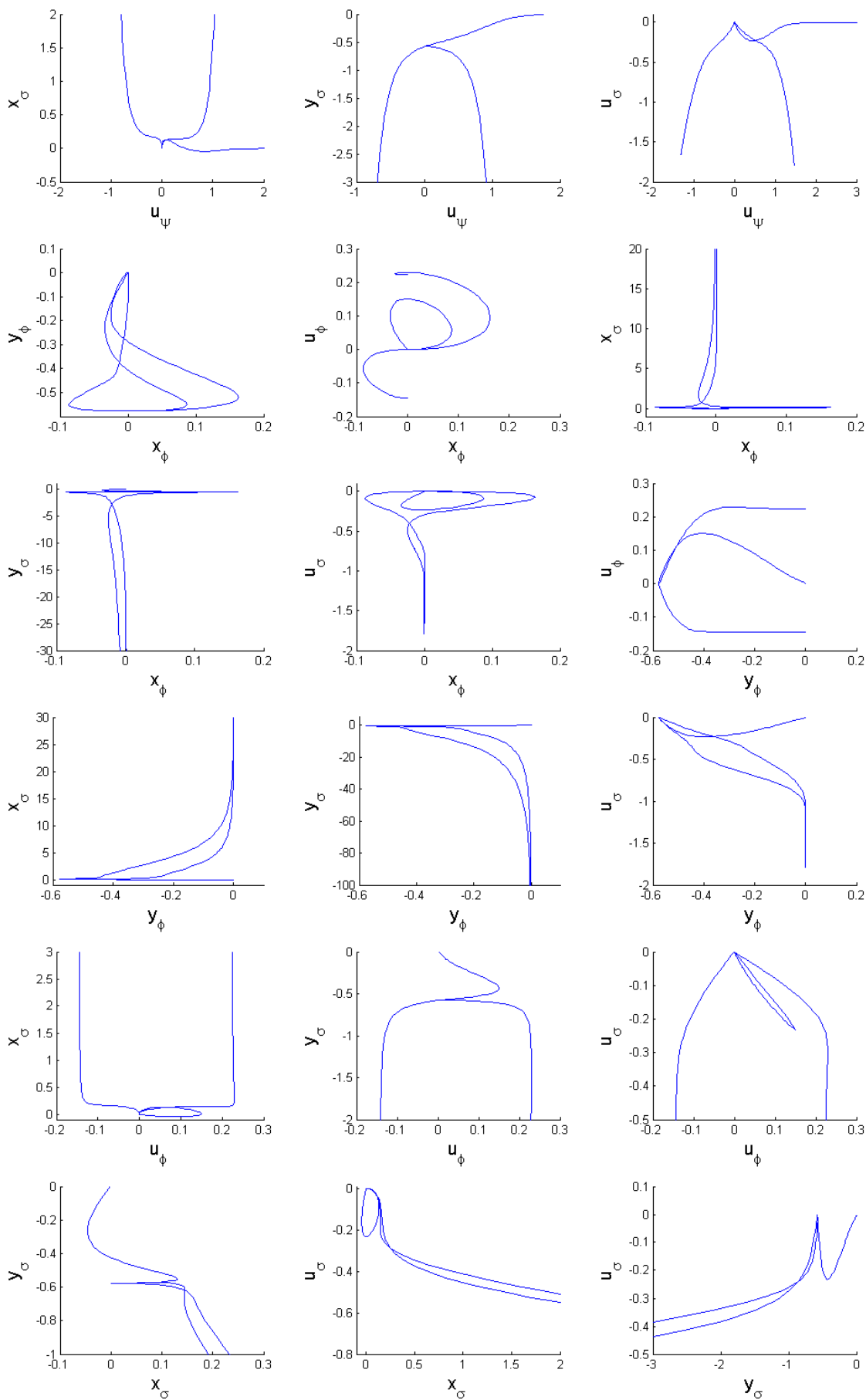

FIGURE 3: Two dimensional projections of the phase-space trajectories for $\xi_1 = \xi_2 = \xi_3 = -0.5$, $\alpha_\psi = \alpha_\phi = \alpha_\sigma = -1.5$. All plots begin from the critical point $C$ being a stable attractor.

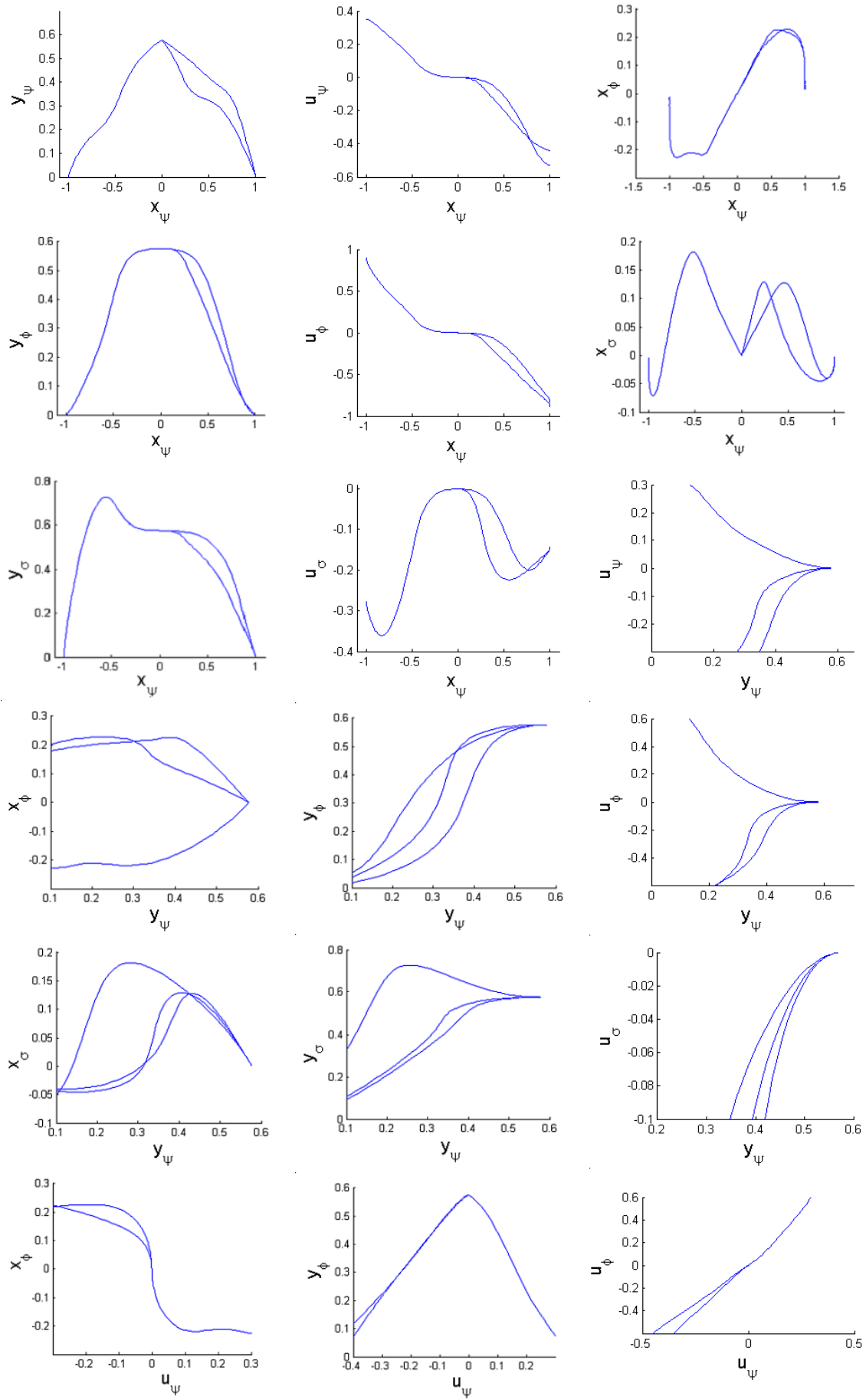

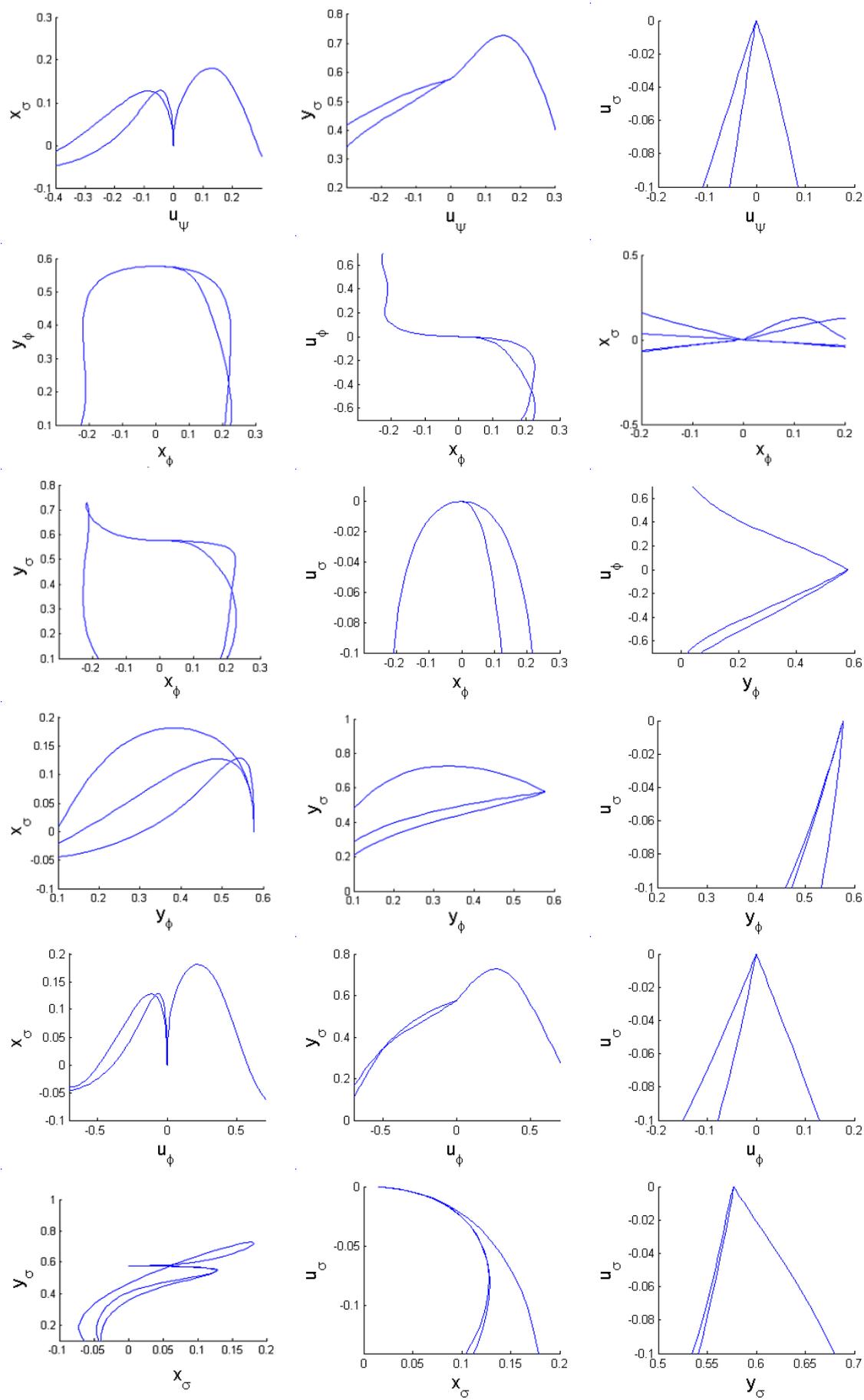

FIGURE 4: Two dimensional projections of the phase-space trajectories for $\xi_1 = \xi_2 = \xi_3 = -0.5$, $\alpha_\psi = \alpha_\phi = \alpha_\sigma = -1.5$. All plots begin from the critical point $D$ being a stable attractor.

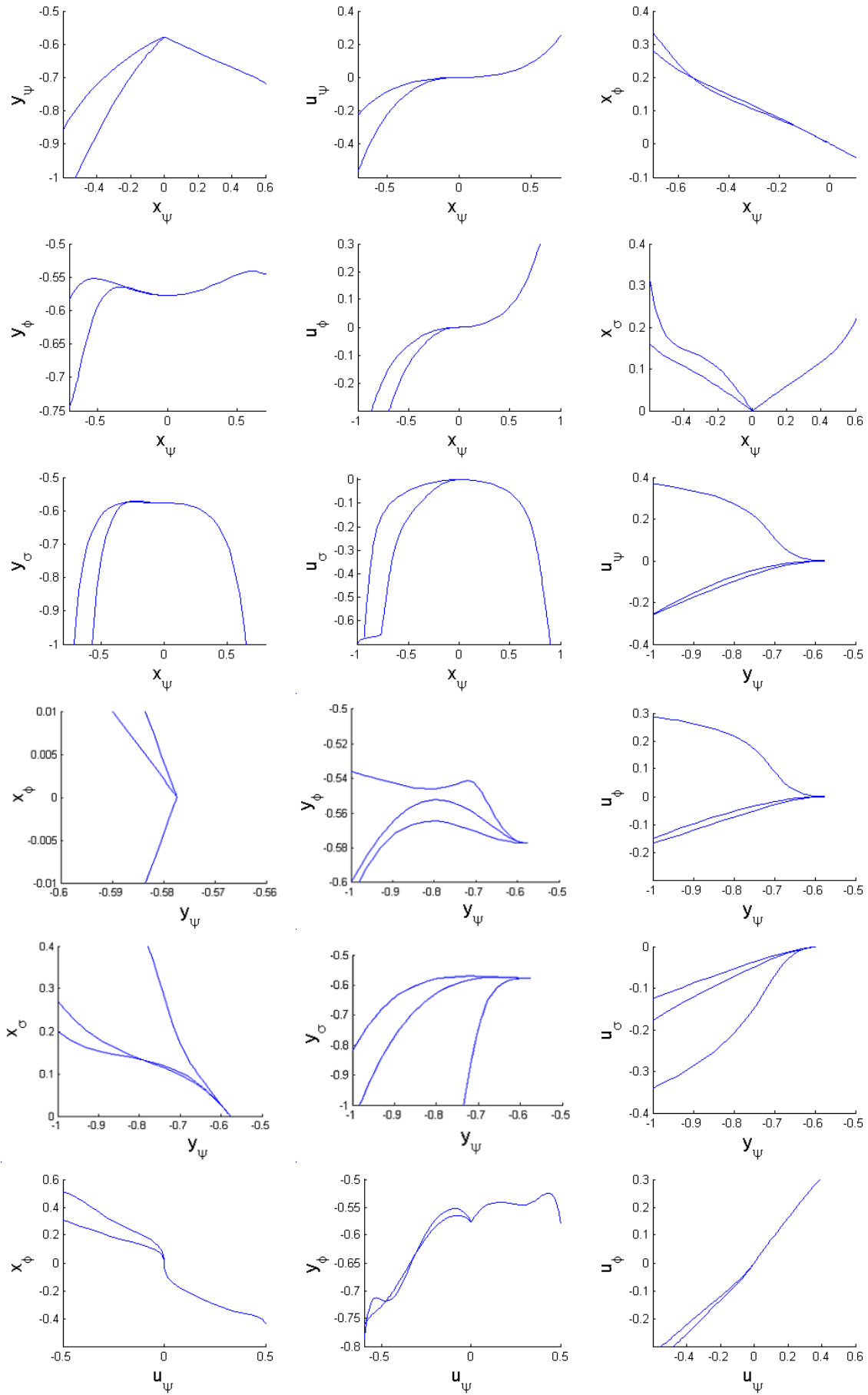

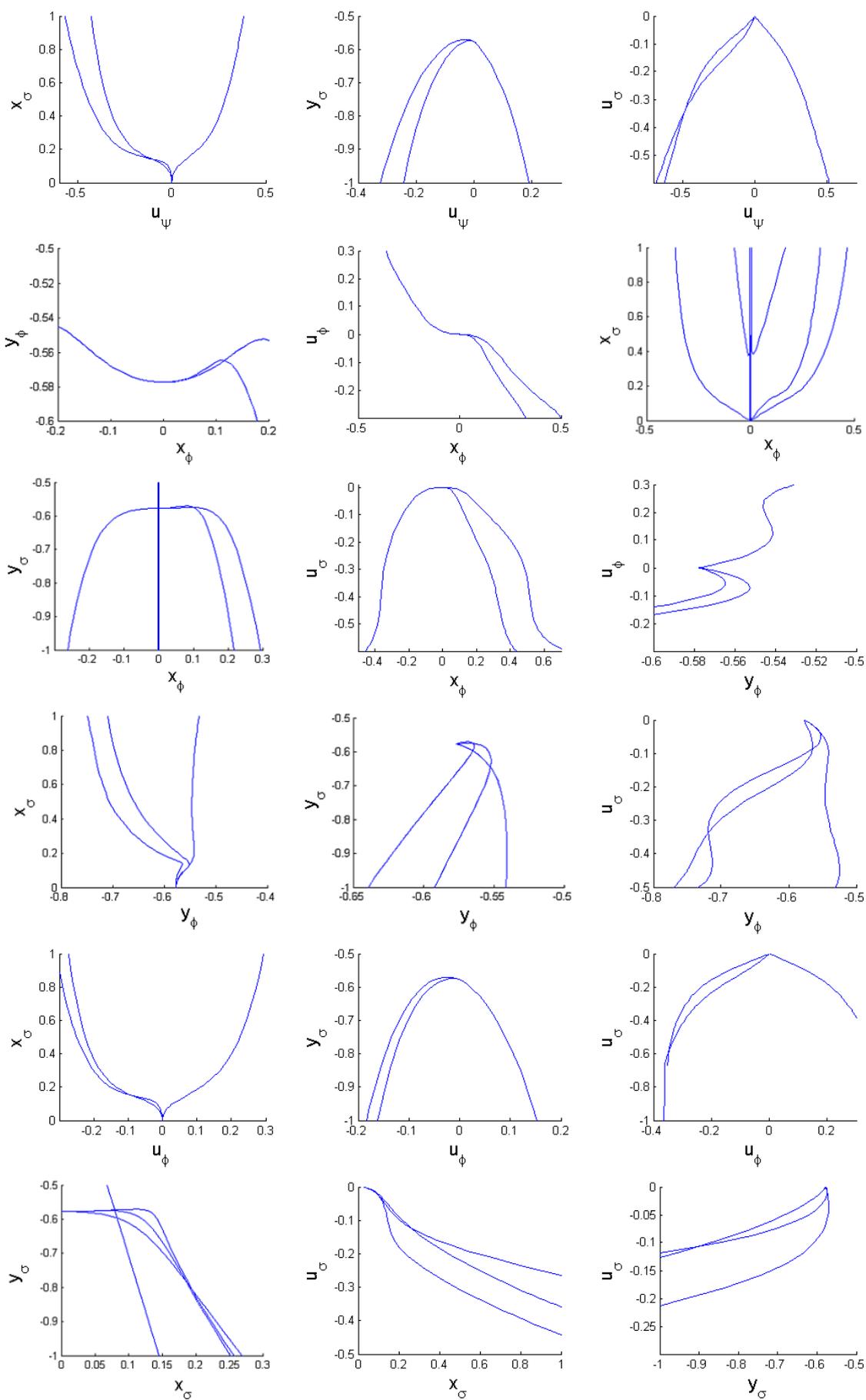